\documentclass[aps,12pt]{revtex4-1}
\usepackage[english]{babel}
\usepackage{amsmath,graphicx,bbm,accents}

\newcommand{\bignone}{}
\newcommand{\mathd}{\mathrm{d}}
\newcommand{\mathe}{\mathrm{e}}
\newcommand{\nocomma}{}
\newcommand{\nosymbol}{}
\newcommand{\tmdate}[1]{\today}
\newcommand{\tmem}[1]{{\em #1\/}}
\newcommand{\tmop}[1]{\ensuremath{\operatorname{#1}}}
\newcommand{\tmsep}{, }
\newcommand{\tmtextbf}[1]{{\bfseries{#1}}}

\makeatletter
\DeclareRobustCommand{\cev}[1]{%
  \mathpalette\do@cev{#1}%
}
\newcommand{\do@cev}[2]{%
  \fix@cev{#1}{+}%
  \reflectbox{$\m@th#1\vec{\reflectbox{$\fix@cev{#1}{-}\m@th#1#2\fix@cev{#1}{+}$}}$}%
  \fix@cev{#1}{-}%
}
\newcommand{\fix@cev}[2]{%
  \ifx#1\displaystyle
    \mkern#23mu
  \else
    \ifx#1\textstyle
      \mkern#23mu
    \else
      \ifx#1\scriptstyle
        \mkern#22mu
      \else
        \mkern#22mu
      \fi
    \fi
  \fi
}

\makeatother
\let \check=\cev

\begin{document}

\title{Quantum renewal processes}

\author{Bassano Vacchini}
\affiliation{Dipartimento di Fisica ``Aldo Pontremoli'', Universit{\`a} degli
Studi di Milano, Via Celoria 16, I-20133 Milan, Italy}
\affiliation{INFN, Sezione di Milano, Via Celoria 16, I-20133 Milan, Italy}

\begin{abstract}
  We introduce a general construction of master equations with memory kernel
  whose solutions are given by completely positive trace preserving maps.
  These dynamics going beyond the Lindblad paradigm are obtained with
  reference to classical renewal processes, so that they are termed quantum
  renewal processes. They can be described by means of semigroup dynamics
  interrupted by jumps, separated by independently distributed time intervals,
  following suitable waiting time distributions. In this framework one can
  further introduce modified processes, in which the first few events follow
  different distributions. A crucial role, marking an important difference
  with respect to the classical case, is played by operator ordering. Indeed,
  for the same choice of basic quantum transformations different quantum
  dynamics arise. In particular for the case of modified processes it is
  natural to consider the time inverted operator ordering, in which the last
  few events are distributed differently.
\end{abstract}


{\maketitle}

\section{Introduction}

The proper description of the dynamics of a quantum system in many cases of
relevance calls for taking into account all other degrees of freedom,
typically called environmental, which might affect its time evolution. In such
cases one speaks of the dynamics of an open quantum system
{\cite{Breuer2002,Rivas2012}}. Indeed, closed systems, strictly isolated from
any other degree of freedom over any time scale, are rather an exception. When
dealing with open quantum systems, a generally valid evolution equation such
as the Schr{\"o}dinger equation for isolated systems is not known. A class of
dynamics which has proven to be of great relevance is given by semigroups,
which break in a natural way the reversibility inherent in the unitary
evolution. These semigroup evolutions are obtained as solution of master
equations whose structure has been fully characterized
{\cite{Lindblad1976a,Gorini1976a}} and is typically called Lindblad form. They
provide the natural quantum counterpart of classical Markovian semigroups, and
indeed has been first introduced in view of this analogy
{\cite{Kossakowski1972a}}. As a result evolutions of Lindblad type has proven
a reference result for all situations in which a Markovian approach can be
considered, and memory effects can be neglected. This is however often not the
case, e.g. due to strong coupling or low temperatures. The characterization of
more general evolution equations, which might take into account non-Markovian
effects, is therefore a pressing issue. In this direction one can consider two
main possible approaches, i.e. either time-local master equations or
integro-differential ones involving a MK (MK). The key difficulty in both
approaches, which has not yet found a general solution, is determining the
conditions on the structure of the master equation warranting trace
preservation and complete positivity of the solutions. Memory kernels
warranting this property are usually termed legitimate. Various efforts have
been done in this direction, leading to partial results both with reference to
equations in time-local form
{\cite{Chruscinski2010a,Breuer2012a,Chruscinski2014a,Andersson2014a,Amato2019a,Bernal2019a,Reimer-xxx}},
as well as to equations in time non local form
{\cite{Barnett2001a,Daffer2004a,Budini2004a,Shabani2005a,Breuer2008a,Wilkie2009a,Campbell2012a,Ciccarello2013a,Budini2013b,Vacchini2013a,Chruscinski2016a,Vacchini2016b}}.

In this paper we will provide a derivation of classes of legitimate MK,
relying on the analogy with classical stochastic processes. Obtaining MK
master equations has shown to be a daunting task, but it appears that very
large classes can be introduced and connected to a very simple physical
interpretation as well as a natural probabilistic interpretation. The quantum
processes arising as solution of these equations are connected to quantum
versions of classical renewal processes and modified renewal processes, and
are characterized by the fact that they provide a piecewise continuous
dynamics in which continuous time evolutions of semigroup type are interrupted
by jumps described as completely positive trace preserving (CPT)
transformations. These jumps are distributed in time according to waiting time
distributions (WTD) appearing in the characterization of renewal processes,
including modified renewal processes, in which the first few time intervals
are different from the following ones. The starting point of this analysis
will be a suitable correspondence rule from classical commuting quantities to
operators, in the same spirit of {\cite{Chruscinski2016a,Vacchini2016b}}. Two
new aspects are considered for the first time in this work, thus allowing to
significantly enlarge the class of known quantum MK warranting as solutions
legitimate dynamics: the introduction of modified quantum renewal processes
and the consideration of inverse time operator ordering, which still leads to
well-defined dynamics. This approach encompasses simple examples already
considered in the literature
{\cite{Herzog1995a,Cresser1996a,Budini2004a,Ciccarello2013a,Lorenzo2016a,Cresser2019a}}
and puts them within a more general theory. It thus opens the way for
considering more general dynamics, e.g. in the framework of collision models,
which have recently attracted a lot of interest providing a powerful tool to
address issues in quantum thermometry, quantum thermodynamics, quantum optics,
quantum entanglement and quantum non-Markovianity
{\cite{Paternostro2014a,Bernardes2015a,Ciccarello2017a,Strasberg2017a,Campbell2018a,Jin2018a,Breuer2018a,Cakmak2019a,Campbell2019a,Seah-xxx,Shaji-xxx}}.
Indeed, collision models are naturally introduced as dynamics characterized by
a sequence of collisions or jumps. The variety of such models in the
dependence on the jump operators as well as features and possible interactions
between the environmental components has been extensively analyzed
{\cite{Ziman2005a,Rybar2012a,Paternostro2014a,Kretschmer2016a,Baris2017a,Jin2018a,Shaji-xxx}},
while little has been done to investigate the relevance of the distribution in
time of the interaction events. This theoretical proposal provides a
groundwork for the study of these effects, allowing in particular to deal with
situations in which a selection of collision events have to be treated
differently.

The paper is organized as follows. In Sec.~\ref{s1} we discuss previous
approaches to the introduction of MK, while in Sec.~\ref{s3} we introduce the
notion of quantum renewal process. In Sec.~\ref{s4} and ~\ref{s5} we
investigate the different processes arising by considering modified WTD and
the inverse time order in the allocation of jumps respectively. In
Sec.~\ref{s6} we consider a few simple examples, finally pointing to possible
developments in Sec.~\ref{s7}.

\section{Memory kernels and generalized master equations}\label{s1}

Let us first recall the general framework. We say that the dynamics of a
system is described by a MK master equation if the time dependent statistical
operator $\rho ( t )$ describing the statistics of observations on the system
obeys
\begin{eqnarray}
  \frac{\mathd}{\mathd t} \rho ( t ) & = & \! \int^{t}_{0} \mathd \tau
  \mathcal{W} ( t- \tau ) \rho ( \tau ) + \mathcal{I} ( t ) \rho ( 0 ) ,
  \label{eq:wi} 
\end{eqnarray}
where the superoperator $\mathcal{W} ( t )$ is called MK, while the
superoperator $\mathcal{I} ( t )$ is usually termed inhomogeneous
contribution. We stress the fact that the term memory is used because one is
faced with an integral equation with respect to the operator-valued variable
$\rho ( t )$. This is not directly related to a notion of memory in the
quantum dynamics, as possibly captured by the different recently introduced
notions of quantum non-Markovianity
{\cite{Rivas2014a,Breuer2016a,Devega2017a}}. Since the master equation
Eq.~(\ref{eq:wi}) is meant to describe the evolution in time of a statistical
operator, the corresponding solutions should comply with two basic
requirements, namely preservation of trace and positivity of the state.
Assuming that this dynamics arises as a consequence of the interaction of the
system of interest with some environment, then also complete positivity has to
be asked for. Introducing the linear transformation $\Lambda ( t )$ giving the
time evolution
\begin{eqnarray}
  \rho ( t ) & = & \Lambda ( t ) \rho ( 0 ) , \nonumber
\end{eqnarray}
and therefore obeying
\begin{eqnarray}
  \frac{\mathd}{\mathd t} \Lambda ( t ) & = & \! \int^{t}_{0} \mathd \tau
  \mathcal{W} ( t- \tau ) \Lambda ( \tau ) + \mathcal{I} ( t ) \label{eq:l} 
\end{eqnarray}
with the initial condition $\Lambda ( 0 ) = \mathbbm{1}$, these requirements
correspond to take $\{ \Lambda ( t ) \}_{t \in \mathbbm{R}_{+}}$ as a
collection of CPT maps.

The quest for introducing MK, and possibly corresponding inhomogeneous
contributions, which lead to well-defined quantum transformation going beyond
the standard Lindblad dynamics, has proven to be quite hard, though some
reference results have been obtained, formulating either sufficient or
necessary conditions on the superoperator expressions. In particular, making
reference to the theory of semi-Markov processes, a class of non-Markovian
classical stochastic processes, it has proven possible to obtain a large
collection of legitimate quantum MK. To obtain such dynamics, which have been
termed quantum semi-Markov processes, one considers a quantum master equation
of the form Eq.~(\ref{eq:l}) with a vanishing inhomogeneous term, namely
\begin{eqnarray}
  \frac{\mathd}{\mathd t} \rho ( t ) & = & \! \int^{t}_{0} \mathd \tau
  \mathcal{K} ( t- \tau ) \rho ( \tau ) , \label{eq:wik} 
\end{eqnarray}
with MK superoperators given by $\mathcal{\widehat{\mathcal{K}}}_{l} ( u ) =u+
\widehat{g \mathcal{G}} ( u )^{-1} ( \widehat{f \mathcal{F}} ( u ) -
\mathbbm{1} )$ and $\mathcal{\widehat{\mathcal{K}}}_{r} ( u ) = u+ (
\widehat{f \mathcal{F}} ( u ) - \mathbbm{1} ) \widehat{g \mathcal{G}} ( u
)^{-1}$, where the indexes $l,r$ denote left and right respectively, in view
of operator ordering. The kernels are built in terms of the Laplace transform
of the operators $f ( t ) \mathcal{F} ( t )$ and $g ( t ) \mathcal{G} ( t )$,
where $f ( t )$ is a WTD, and $g ( t )$ the corresponding survival
probability. These MK master equations have solutions given by CPT
transformations if $\{ \mathcal{F} ( t ) \}_{t \in \mathbbm{R_{+}}}$ and $\{
\mathcal{G} ( t ) \}_{t \in \mathbbm{R_{+}}}$ are arbitrary collection of CPT
maps, with the only further constraint $\mathcal{G} ( 0 ) = \mathbbm{1}$. The
associated time evolutions can be shown to be given by the collections of maps
\begin{eqnarray}
  \hat{\Lambda}_{l} ( u ) & = & \widehat{g \mathcal{G}} ( u ) ( \mathbbm{1} -
  \widehat{f \mathcal{F}} ( u ) )^{-1}  \label{eq:ll}
\end{eqnarray}
and
\begin{eqnarray}
  \hat{\Lambda}_{r} ( u ) & = & ( \mathbbm{1} - \widehat{f \mathcal{F}} ( u )
  )^{-1} \widehat{g \mathcal{G}} ( u ) ,  \label{eq:lr}
\end{eqnarray}
respectively. Though different approaches have been considered to obtain MK
falling in this class {\cite{Budini2004a,Chruscinski2016a,Chruscinski2017a}},
the possibly simplest starting point to recover and understand these results
is to make contact with the generalized master equation for the transition
probability $T_{nm} (t)$ of a semi-Markov process
{\cite{Feller1964a,Gillespie1977a,Nollau1980,Breuer2009a}} which reads
\begin{equation}
  \frac{\mathd}{\tmop{dt}} T_{nm} (t) = \int_{0}^{t} d \tau \sum_{k} [ w_{nk}
  ( t- \tau ) T_{km} ( \tau )-w_{kn} ( t- \tau ) T_{nm} ( \tau ) ] ,
  \label{eq:start}
\end{equation}
where $T_{nm} (t)$ provides the probability to reach site $n$ at time $t$
given that one starts from an arbitrary but fixed site $m$ at time $t=0$.
Indeed, a semi-Markov process describes the time evolution of a classical
system which can jump among different sites according to fixed probabilities,
the time elapsing between subsequent jumps being described by a collection of
independent and identically distributed random variables, which might depend
on the considered site. The process is specified by a so-called semi-Markov
matrix, a time dependent matrix whose entries provide the probability density
to jump between two sites in a given time. For the special case in which the
semi-Markov matrix is given by a stochastic matrix times an exponential WTD,
one recovers a classical Markovian jump process. In all other cases the
classical process is non-Markovian. The semi-Markov matrix determines the MK
$w_{nk} ( t )$ appearing in Eq.~(\ref{eq:start}), which in Laplace transform
reads $\hat{w}_{nk} (u) = \hat{g}_{n} (u) \pi_{nk} \hat{f}_{k} (u) /
\hat{g}_{k} (u)$. Here $\pi_{nk}$ are the elements of the stochastic matrix
whose entries are the jump probabilities between sites, while $f_{n} ( t )$
provides the WTD at site $n$, namely the probability distribution for the time
elapsing before the next jump takes place. The function $g_{n} ( t )$ is the
corresponding survival probability, given by $g_{n} ( t ) =1- \int_{0}^{t} d
\tau f_{n} ( \tau )$. In Laplace transform the solution of the generalized
master equation Eq.~(\ref{eq:start}) reads
\begin{eqnarray}
  \hat{T}_{nm} (u) & = & \sum_{k} \bignone ( \mathbbm{1} - \pi \hat{f}_{} (u)
  )_{nk}^{-1} \hat{g}_{k} (u) \hat{T}_{km} (0) . \label{eq:csol} 
\end{eqnarray}
The quantum maps Eq.~(\ref{eq:ll}), Eq.~(\ref{eq:lr}) and related quantum MK
$\widehat{\mathcal{K}}_{l,r} ( u )$ can thus be obtained by using the
following correspondence rule between functions and operator-valued
expressions
\[ f ( t ) \rightarrow f ( t ) \mathcal{F} ( t ) , \hspace{1em} g ( t )
   \rightarrow g ( t ) \mathcal{G} ( t ) \]
together with a choice of operator ordering, which we have encoded in the
$l,r$ index. The relevance of operator ordering, rooted in non commutativity
of quantum transformations, brings with itself the fact that the very same
classical kernel can lead to different quantum kernels. The maps $\mathcal{F}
( t )$ and $\mathcal{G} ( t )$ describe the time evolution of the quantum
system in between jumps, and the effect of the stochastic matrix $\pi$, which
is naturally replaced by a CPT map, has been reabsorbed in the collection $\{
\mathcal{F} ( t ) \}_{t \in \mathbbm{R_{+}}}$, which upon composition with a
fixed CPT map remains in the class.

In this framework one has two basic results. On the one hand one obtains a
characterization of a very large class of legitimate quantum kernels; on the
other hand the resulting dynamics can be naturally described as a piecewise
quantum dynamics, in which continuous in time quantum evolutions are reset at
given times or interrupted by jumps.

The obtained dynamics, for specific choices of the involved collections of
maps and CPT transformations, has been shown to be connected to physical
models. More specifically maps of the form Eq.~(\ref{eq:ll}) provide the
mathematical description for the dynamics of the micromaser
{\cite{Cresser1992a,Raithel1994a,Englert2002b}}, while transformations as in
Eq.~(\ref{eq:lr}) correspond to classes of collision models with memory
{\cite{Giovannetti2012a,Ciccarello2013a,Lorenzo2016a}}.

\section{Quantum renewal processes}\label{s3}

\begin{figure}[h]
  {\includegraphics[width=0.8\textwidth]{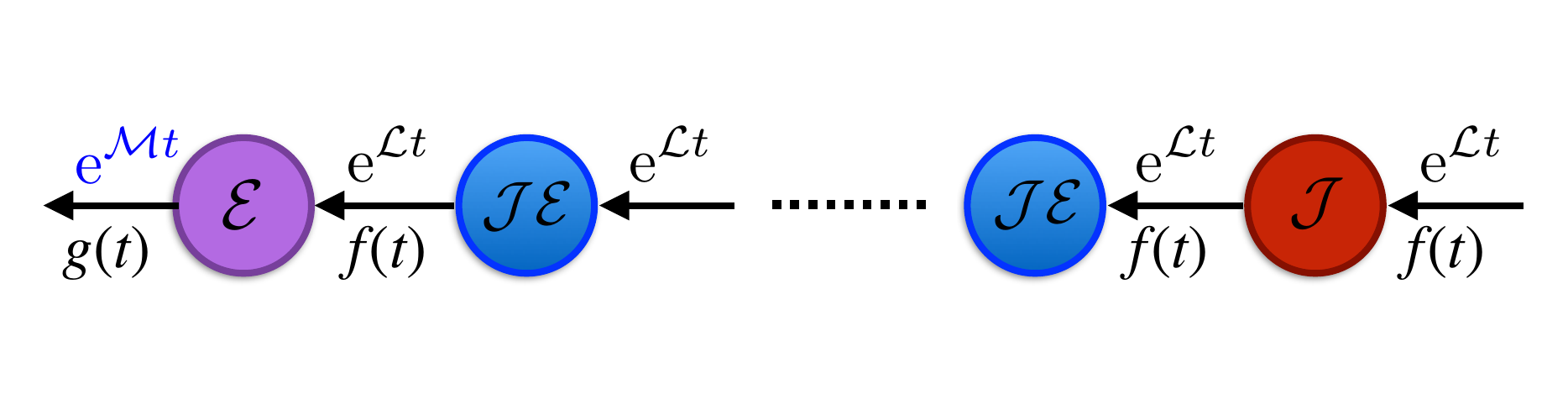}}
  \caption{\label{fig:schema-je}Scheme of a time evolution described by a
  quantum renewal process. In between jumps the time evolution is given by a
  semigroup, possibly a different one for the last time interval. The jumps
  are described by CPT transformations which might differ in both initial and
  final application.}
\end{figure}

We now consider the case in which the collection of CPT maps $\{ \mathcal{F} (
t ) \}_{t \in \mathbbm{R_{+}}}$ and $\{ \mathcal{G} ( t ) \}_{t \in
\mathbbm{R_{+}}}$ can be obtained as quantum dynamical semigroups composed
with fixed jump transformations. In this setting we fix the dynamics taking
place in between jumps and concentrate on the effect of the jump
transformations and the time elapsed in between jumps, so that by analogy with
classical renewal processes {\cite{Cox1965}} it is natural to call such
dynamics quantum renewal processes. We therefore take the collections $\{
\mathcal{F} ( t ) \}_{t \in \mathbbm{R_{+}}}$ of maps to be of the form
\begin{eqnarray}
  \mathcal{F} ( t ) & = & \mathcal{E} \mathe^{\mathcal{L} t} \mathcal{J} ,
  \label{eq:expl} 
\end{eqnarray}
with $\mathcal{L}$ an arbitrary generator in Lindblad form, while
$\mathcal{E}$ and $\mathcal{J}$ are arbitrary CPT maps, together with
$\mathcal{G} ( t ) = \mathe^{\mathcal{M} t}$, according to the relation
$\mathcal{G} ( 0 ) = \mathbbm{1}$. For each of these choices of time dependent
transformations we have two distinct kernels, arising due to $l$ and $r$
operator ordering. Focussing on the $l$ case, exploiting the fact that
multiplication by an exponential function in Laplace transform goes over to
translation, as shown in Appendix~\ref{a1} we have the kernel
\begin{eqnarray}
  \widehat{\mathcal{K}}_{l} ( u ) & = & \mathcal{M} + ( \mathcal{E} \hat{f}
  (u- \mathcal{L} ) \mathcal{J} - \hat{f} (u- \mathcal{M} ) ) \hat{g} ( u-
  \mathcal{M} )^{-1} . \label{eq:kerl} \nonumber
\end{eqnarray}
Considering the corresponding expression of the evolution map
$\hat{\Lambda}_{l} ( u ) = \hat{g} ( u- \mathcal{M} ) ( \mathbbm{1} -
\mathcal{E} \hat{f} (u- \mathcal{L} ) \mathcal{J} )^{-1}$ in time, so that
multiplication goes over to convolution, expanding the Neumann series we
therefore obtain
\begin{eqnarray}
  \rho ( t ) & = & g ( t ) \mathe^{\mathcal{M} t} \rho ( 0 ) +
  \sum_{n=1}^{\infty} \int^{t}_{0} \mathd t_{n} \ldots \int^{t_{2}}_{0} \mathd
  t_{1}  \\
  &  & \times g (t-t_{n} ) \mathe^{\mathcal{M}  (t-t_{n} )} \mathcal{E} f
  (t_{n} -t_{n-1} ) \mathe^{\mathcal{L}  (t_{n} -t_{n-1} )} \ldots \mathcal{J}
  \mathcal{E} f ( t_{1} ) \mathe^{\mathcal{L} t_{1}} \mathcal{J} \rho ( 0 )
  \label{eq:je} . \nonumber
\end{eqnarray}
The evolution is thus described as a piecewise dynamics, interrupted by jumps,
in which initial, final and intermediate transformations can be different, as
shown in Fig.~\ref{fig:schema-je}.

At the same time trace preservation is generally warranted by the fact that
\begin{eqnarray}
  g ( t_{} ) + \sum_{n=1}^{\infty} \int^{t}_{0} \mathd t_{n} \ldots
  \int^{t_{2}}_{0} \mathd t_{1} g (t-t_{n} ) f (t_{n} -t_{n-1} ) \ldots f
  (t_{2} -t_{1} ) f ( t_{1} ) & = & 1 \label{eq:prob} 
\end{eqnarray}
as follows from the theory of renewal processes {\cite{Cox1965}}. In
particular one can consider a different semigroup evolution for the time
before the last jump. All these dynamics share the fact of being describable
as a combination of semigroup dynamics over independent identically
distributed time intervals. Non-commutativity implies that at variance with
the classical case, for quantum renewal processes also the time at which the
jumps take place affects the dynamics. Special examples of this framework has
been previously considered in the literature, as one of the first examples of
legitimate MK {\cite{Budini2004a,Budini2005a}}. While we stress the fact that
even in the simplified case in which one of the transformations is trivial,
i.e. either $\mathcal{E}$ or $\mathcal{J}$ is the identity transformation,
Eq.~(\ref{eq:expl}) combined with the choice of ordering leads to four
distinct quantum dynamics, it is of interest to work out in more detail a
special case, to show the connection with the standard Markovian semigroup
dynamics. To determine the dynamics we have to specify different quantities,
namely the generators $\mathcal{L}$ and $\mathcal{M}$, the quantum channels
$\mathcal{E}$ and $\mathcal{J}$, as well as the WTD $f ( t )$. Let us take
$\mathcal{M} = \mathcal{L}$, $\mathcal{J} = \mathbbm{1}$ and $f ( t ) =
\lambda \mathe^{- \lambda t}$, that is we consider an exponential waiting
time, which in the classical case leads to a Markov renewal process, namely a
Poisson process. As shown in Appendix~\ref{a1} the MK takes the form
$\mathcal{K} ( t ) = \delta ( t ) [ \mathcal{L} + \lambda ( \mathcal{E} -
\mathbbm{1} ) ]$, corresponding to a semigroup dynamics given by the sum of
two generators
\begin{eqnarray}
  \Lambda ( t ) & = & \mathe^{[ \mathcal{L} + \lambda ( \mathcal{E} -
  \mathbbm{1} ) ] t} . \label{eq:split} 
\end{eqnarray}
One can now exploit the relation $( u- ( A+B ) )^{-1} = ( u-A )^{-1} ( 1-B (
u-A )^{-1} )^{-1}$ valid for two arbitrary operators $A$ and $B$ and leading
to the Dyson expansion
\begin{eqnarray}
  \mathe^{( A+B ) t} & = & \mathe^{At} + \sum_{n=1}^{\infty} \int^{t}_{0}
  \mathd t_{n} \ldots \int^{t_{2}}_{0} \mathd t_{1} \mathe^{A ( t-t_{n} )} B
  \mathe^{A ( t_{n} -t_{n-1} )} \ldots B \mathe^{At_{1}} , \label{eq:Dyson} 
\end{eqnarray}
for the two possible splitting of the argument of the exponential in
Eq.~(\ref{eq:split}). The apparently most natural choice is $A= \mathcal{L}$
and $B= \lambda ( \mathcal{E} - \mathbbm{1} )$, leading to
\begin{eqnarray}
  \Lambda (t) & = & \mathe^{\mathcal{L} t} + \sum_{n=1}^{\infty} \int^{t}_{0}
  \mathd t_{n} \ldots \int^{t_{2}}_{0} \mathd t_{1} \times \nonumber\\
  &  & \mathe^{\mathcal{L}  (t-t_{n} )} \lambda ( \mathcal{E} - \mathbbm{1} )
  \ldots \mathe^{\mathcal{L}  (t_{2} -t_{1} )} \lambda ( \mathcal{E} -
  \mathbbm{1} ) \mathe^{\mathcal{L} t_{1}} \label{eq:DysonS} 
\end{eqnarray}
to be compared with the alternative choice $A=\mathcal{E}- \lambda
\mathbbm{1}$ and $B= \lambda \mathcal{E}$ leading to
\begin{eqnarray}
  \Lambda (t) & = & \mathe^{- \lambda t} \mathe^{\mathcal{L} t} +
  \sum_{n=1}^{\infty} \int^{t}_{0} \mathd t_{n} \ldots \int^{t_{2}}_{0} \mathd
  t_{1} \times \nonumber\\
  &  & \mathe^{- \lambda (t_{} -t_{n} )} \mathe^{\mathcal{L}  (t-t_{n} )}
  \mathcal{E} \ldots \mathcal{E} \lambda \mathe^{- \lambda (t_{2} -t_{1} )}
  \mathe^{\mathcal{L}  (t_{2} -t_{1} )} \mathcal{E} \lambda \mathe^{- \lambda
  t_{1}} \mathe^{\mathcal{L} t_{1}} . \label{eq:Dysonf} 
\end{eqnarray}
Both representations are exact. It now immediately appears that
Eq.~(\ref{eq:Dysonf}), arising from a mixture with positive coefficients of
Lindblad generators, is a special case of Eq.~(\ref{eq:je}) for the choice of
an exponential WTD with rate $\lambda$, together with $\mathcal{J} =
\mathbbm{1}$ and $\mathcal{M} = \mathcal{L}$. Also the equivalent expression
Eq.~(\ref{eq:DysonS}) can be written in a way which allows to connect to a
generic WTD. Indeed, a renewal process is uniquely determined from its WTD or
equivalently its renewal density $S ( t )$, also known as sprinkling
distribution, arising as solution of the renewal equation $S ( t ) =f ( t ) +
\int^{t}_{0} \mathd \tau f ( t- \tau ) S ( \tau ) \bignone$. For the case of a
memoryless exponential waiting time the sprinkling distribution, which gives
the probability density to have a jump at the given time, neglecting all
previous jumps, is a constant function, simply given by the rate $\lambda$.
Indeed, one can check that for $\mathcal{J} = \mathbbm{1}$ and $\mathcal{M} =
\mathcal{L}$ the original time evolution Eq.~(\ref{eq:je}) allows for the two
equivalent expressions
\begin{eqnarray}
  \rho ( t ) & = & g ( t ) \mathe^{\mathcal{L} t} \rho ( 0 ) +
  \sum_{n=1}^{\infty} \int^{t}_{0} \mathd t_{n} \ldots \int^{t_{2}}_{0} \mathd
  t_{1}  \nonumber\\
  &  & \times g (t-t_{n} ) \mathe^{\mathcal{L}  (t-t_{n} )} \mathcal{E} f
  (t_{n} -t_{n-1} ) \mathe^{\mathcal{L}  (t_{n} -t_{n-1} )} \ldots \mathcal{E}
  f ( t_{1} ) \mathe^{\mathcal{L} t_{1}} \rho ( 0 ) \label{eq:ef} \\
  & = & \mathe^{\mathcal{L} t} \rho ( 0 ) + \sum_{n=1}^{\infty} \int^{t}_{0}
  \mathd t_{n} \ldots \int^{t_{2}}_{0} \mathd t_{1}  \nonumber\\
  &  & \times \mathe^{\mathcal{L}  (t-t_{n} )} ( \mathcal{E} - \mathbbm{1} )
  S (t_{n} -t_{n-1} ) \mathe^{\mathcal{L}  (t_{n} -t_{n-1} )} \ldots (
  \mathcal{E} - \mathbbm{1} ) S (t_{1} ) \mathe^{\mathcal{L} t_{1}} \rho ( 0 )
  \label{eq:es} , 
\end{eqnarray}
as follows from the operator identity
\begin{eqnarray}
  \hat{g} ( u- \mathcal{\mathcal{L}} ) \frac{1}{\mathbbm{1} - \mathcal{E}
  \hat{f} (u- \mathcal{L} )} & = & \frac{1}{u- \mathcal{L}} 
  \frac{1}{\mathbbm{1} - ( \mathcal{E} - \mathbbm{1} ) \hat{S} (u- \mathcal{L}
  )} , \label{eq:fs} 
\end{eqnarray}
proven in Appendix.~\ref{a2}.

The expression Eq.~(\ref{eq:ef}) of the time evolved state can be interpreted
as a sum of contributions corresponding to a piecewise dynamics with a
different number of intermediate jumps. During the jumps described by the CPT
map $\mathcal{E}$ the state evolves according to a semigroup dynamics
determined by $\mathcal{L}$ for a time interval $(t_{n} -t_{n-1} )$ fixed by
the waiting time $f (t_{n} -t_{n-1} )$. Each term in the sum provides a
contribution to the trace of $\rho ( t )$, corresponding to subcollections
characterized by a given number of jumps. In a complementary way expression
Eq.~(\ref{eq:es}) is the sum of a purely semigroup dynamics together with
terms determined by the repeated appearance of contributions of the form $(
\mathcal{E} - \mathbbm{1} ) S (t_{n} -t_{n-1} )$. The latter can be
interpreted saying that with a probability density given by the sprinkling
distribution $S (t_{n} -t_{n-1} )$ the time evolved contribution is replaced
by another in which an additional $\mathcal{E}$ transformation has acted upon,
hence the operator $( \mathcal{E} - \mathbbm{1} )$. In between these
transformations one still has a semigroup dynamics.

The different MK and related time evolutions considered above differ by the
choice of generators $\mathcal{L}$ and $\mathcal{M}$, the choice of channels
$\mathcal{E}$ and $\mathcal{J}$, as well as WTD $f ( t )$. The appearance of
$f ( t )$ warrants trace preservation, while details of the dynamics are
determined by the different operators. We have however always made reference
to the kernel $\mathcal{K}_{l} ( t )$ corresponding to one choice of operator
ordering, that is a specific order in time in which events takes place. This
marks an important difference with respect to the classical case, which we
shall put in better evidence considering modified renewal processes.

\section{Modified quantum renewal processes}\label{s4}

We now derive another class of quantum renewal processes, which can be named
modified renewal processes since in analogy with the classical case they
correspond to a situation in which the WTD characterizing the first $k$
intervals differ from the following ones. Starting from the identity
Eq.~(\ref{eq:prob}), which warranted trace preservation in the previous
examples, moving to the Laplace transform and exploiting $u \hat{g} (u) = ( 1-
\hat{f} (u) )$, one obtains
\begin{eqnarray}
  \frac{1}{u} & = & \hat{g}_{1} (u) + \ldots + \hat{g}_{k} ( u ) \hat{f}_{k-1}
  (u) \ldots \hat{f}_{1} (u) + \hat{g} ( u ) \frac{1}{1- \hat{f} (u)}
  \hat{f}_{k} (u) \ldots \hat{f}_{1} (u) \label{eq:gk} , 
\end{eqnarray}
describing the normalization condition for the situation in which the first
$k$ jumps have a different waiting time. Here again $g_{k} ( t )$ denotes the
survival probability associated to the WTD $f_{k} ( t )$ according to $g_{k} (
t ) =1- \int_{0}^{t} d \tau f_{k} ( \tau )$. A quantum dynamics corresponding
to such modified renewal processes can be obtained via the operator
replacements
\[ \hat{f}_{k} ( u ) \rightarrow \hat{f}_{k} ( u- \mathcal{L} ) \nocomma
   \hspace{1em} \hat{g}_{k} ( u ) \rightarrow \hat{g}_{k} ( u- \mathcal{L} ) ,
\]
leading to the collection of CPT maps
\begin{eqnarray}
  \hat{\Lambda}_{\check{k}} (u) & = & \hat{g}_{1} ( u- \mathcal{L} ) + \ldots
  + \hat{g}_{k} ( u- \mathcal{L} ) \mathcal{E} \hat{f}_{k-1} (u- \mathcal{L} )
  \ldots \mathcal{E} \hat{f}_{1} (u- \mathcal{L} ) \nonumber\\
  &  & + \hat{g} ( u- \mathcal{L} ) \frac{1}{\mathbbm{1} - \mathcal{E}
  \hat{f} (u- \mathcal{L} )} \mathcal{E} \hat{f}_{k} (u- \mathcal{L} ) \ldots
  \mathcal{E} \hat{f}_{1} (u- \mathcal{L} ) \nosymbol , \label{eq:lk} 
\end{eqnarray}
where the arrow appearing in the index denotes the natural time order from
right to left in distinguishing the waiting times. The MK associated to these
modified dynamics are quite involved, but it is natural to express them and
the associated evolution equations making reference to the MK for the
unmodified case, and considering the effect of the modified waiting times by
means of inhomogeneous contributions to the equation. We therefore first
introduce the unmodified dynamics
\begin{eqnarray}
  \hat{\Lambda}_{\check{0}} (u) & = & \hat{g} ( u- \mathcal{L} )
  \frac{1}{\mathbbm{1} - \mathcal{E} \hat{f} (u- \mathcal{L} )} ,
  \label{eq:l0l} 
\end{eqnarray}
so that according to Eq.~(\ref{eq:ll}) we have for the related kernel
\begin{eqnarray}
  \widehat{\mathcal{K}}_{\check{0}} (u) & = & \mathcal{L} + ( \mathcal{E} -
  \mathbbm{1} ) \hat{k} (u- \mathcal{L} ) , \label{eq:k0l} 
\end{eqnarray}
where we have introduced the quantity
\begin{eqnarray}
  \hat{k} (u) & = & \frac{\hat{f} (u)}{\hat{g} (u)} , \label{eq:kdef} 
\end{eqnarray}
which corresponds to the classical kernel associated to the renewal process
{\cite{Cox1965}}. Starting from the relation
\begin{eqnarray}
  u \hat{\Lambda}_{\check{k}} (u) - \mathbbm{1} & = &
  \mathcal{\widehat{\mathcal{K}}}_{\check{0}} ( u ) \hat{\Lambda}_{\check{k}}
  (u) + \mathcal{\widehat{\mathcal{I}}}_{\check{k}} ( u ) \nonumber
\end{eqnarray}
we obtain, as shown in Appendix~\ref{a3}
\begin{eqnarray}
  \mathcal{\widehat{\mathcal{I}}}_{\check{k}} ( u ) & = & ( \mathcal{E} -
  \mathbbm{1} ) ( \hat{S}_{1} (u- \mathcal{L} )- \hat{S} (u- \mathcal{L} ) ) +
  \nonumber\\
  &  & ( \mathcal{E} - \mathbbm{1} ) ( \hat{S}_{2} (u- \mathcal{L} )- \hat{S}
  (u- \mathcal{L} ) ) \mathcal{E} \hat{f}_{1} (u- \mathcal{L} ) + \nonumber\\
  &  & \ldots \nonumber\\
  &  & ( \mathcal{E} - \mathbbm{1} ) ( \hat{S}_{k} (u- \mathcal{L} )- \hat{S}
  (u- \mathcal{L} ) ) \mathcal{E} \hat{f}_{k-1} (u- \mathcal{L} ) \ldots
  \mathcal{E} \hat{f}_{1} (u- \mathcal{L} ) , \label{eq:ik} 
\end{eqnarray}
corresponding to the master equation
\begin{eqnarray}
  \frac{\mathd}{\tmop{dt}} \rho ( t ) & = & \mathcal{L} [ \rho ( t ) ] +
  \int^{t}_{0} \mathd \tau ( \mathcal{E} - \mathbbm{1} ) \mathe^{\mathcal{L} 
  (t- \tau )} k ( t- \tau ) \rho ( \tau ) + ( \mathcal{E} - \mathbbm{1} )
  \mathe^{\mathcal{L} t} ( S_{1} ( t ) -S ( t ) ) \rho ( 0 ) \nonumber\\
  &  & + \sum_{r=2}^{k} \bignone \int^{t}_{0} \mathd t_{r-1} \ldots
  \int^{t}_{0} \mathd t_{1}  ( \mathcal{E} - \mathbbm{1} ) \mathe^{\mathcal{L}
  (t-t_{r-1} )} ( S_{r} ( t-t_{r-1} ) -S ( t-t_{r-1} ) ) \nonumber\\
  &  & \hspace{1em} \times \mathcal{E} f_{r-1} (t_{r-1} -t_{r-2} )
  \mathe^{\mathcal{L}  (t_{r-1} -t_{r-2} )} \ldots \mathcal{E} f_{1} ( t_{1} )
  \mathe^{\mathcal{L} t_{1}} \rho ( 0 ) . \label{eq:msk} 
\end{eqnarray}
The master equation can also be written in the form Eq.~(\ref{eq:wik}), with a
MK that can be compactly expressed in terms of the inhomogeneous contribution
Eq.~(\ref{eq:ik})
\begin{eqnarray}
  \mathcal{\widehat{\mathcal{K}}}_{\check{k}} ( u ) & = & \mathcal{L} +
  \frac{1}{\mathbbm{1} + \mathcal{\widehat{\mathcal{I}}}_{\check{k}} ( u )} \{
  ( \mathcal{E} - \mathbbm{1} ) \hat{k} (u- \mathcal{L} )+
  \mathcal{\widehat{\mathcal{I}}}_{\check{k}} ( u ) (u- \mathcal{L} ) \}
  \label{eq:kk} . 
\end{eqnarray}
In particular, if only the first time interval is different from the others
one recovers for the kernel the slightly more compact expression
\begin{eqnarray}
  \mathcal{\widehat{\mathcal{K}}}_{\check{1}} ( u ) & = & \mathcal{L} +
  \frac{1}{\mathbbm{1} - ( \mathcal{E} - \mathbbm{1} ) ( \hat{S} (u-
  \mathcal{L} )- \hat{S}_{1} (u- \mathcal{L} ) )} ( \mathcal{E} - \mathbbm{1}
  ) \hat{k}_{1} (u- \mathcal{L} ) \label{eq:k1} . 
\end{eqnarray}
This provides a straightforward generalization of one of the first results
about quantum MK {\cite{Budini2004a,Budini2005a}}, and for $\mathcal{L} =0$,
that it neglecting the intermediate time evolution, leads to an evolution of
the form $\rho ( t ) = \hat{\Lambda}_{\check{1}} (u) [ \rho ( 0 ) ] =
\sum_{n=1}^{\infty} p_{1} ( n,t ) \bignone \mathcal{E}^{n} [ \rho ( 0 ) ]$,
where $p_{1} ( n,t )$ provide the probabilities to have $n$ jumps up to time
$t$ for the modified process. An alternative representation of the master
equation, which can be more easily connected to
{\cite{Budini2004a,Budini2005a}} is obtained by considering the reference
kernel Eq.~(\ref{eq:l0l}) together with the inhomogeneous contribution
\begin{eqnarray}
  \widehat{\mathcal{I}}_{\check{1}} ( u ) & = & ( \mathcal{E} - \mathbbm{1} )
  ( \hat{S}_{1} (u- \mathcal{L} )- \hat{S} (u- \mathcal{L} ) ) \label{eq:I1} 
\end{eqnarray}
confirming the results obtained in {\cite{Vacchini2016b,Cresser2019a}}. The
expression of the master equation Eq.~(\ref{eq:msk}) shows that the
inhomogeneous contribution, due to the presence of different WTD
characterizing the first jumps, is directly dependent on the initial
condition, as in the standard derivation of MK master equations within
projection operator techniques {\cite{Breuer2002}}.

\section{Inverse time operator ordering}\label{s5}

\begin{figure}[h]
{\includegraphics[width=0.8\textwidth]{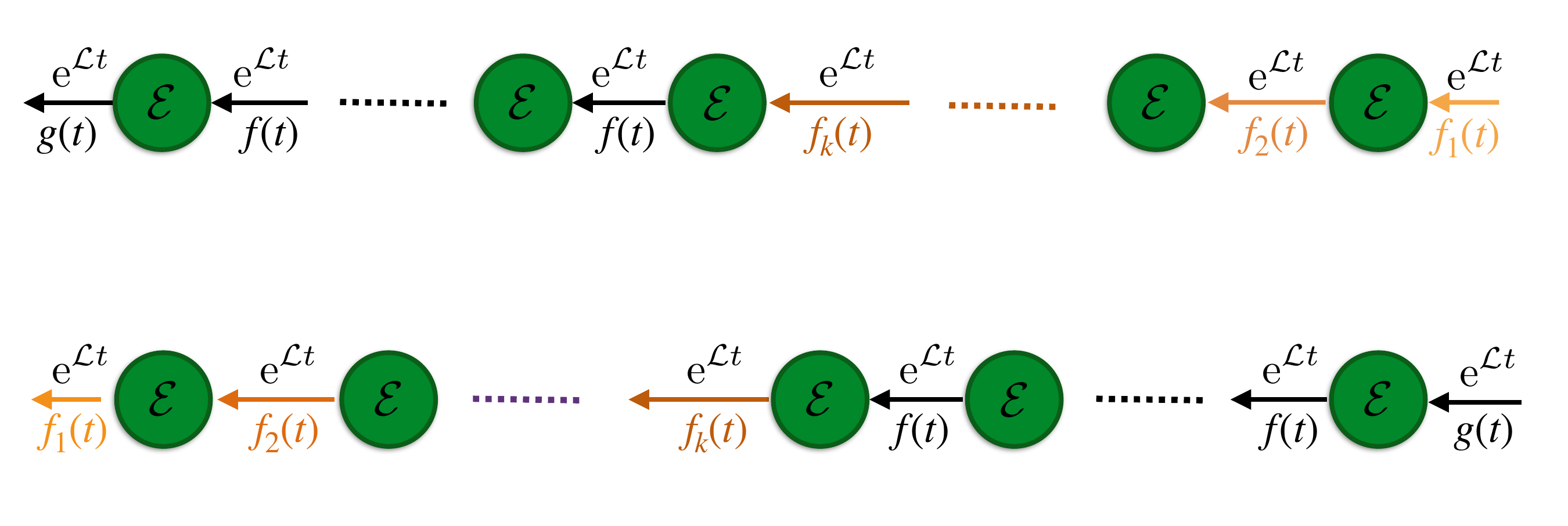}} 
  \caption{\label{fig:schema-lr}Schematic representation of the two possible
  time evolution described by a modified quantum renewal process. In both
  dynamics the jump transformations take place after independent time
  intervals, which are however not all identically distributed. In particular
  one can consider situations in which the first $k$ waiting times (top), or
  the last $k$ ones (bottom) follow different distributions.}
\end{figure}

In the previous analysis we have highlighted the relevance of having non
commuting quantities which, even for a fixed sequence of events, lead to
different evolution equations and different dynamics, at variance with the
classical case. We now put into evidence another peculiar quantum feature,
arising from the fact that when replacing the relation Eq.~(\ref{eq:gk}) with
the operator valued Eq.~(\ref{eq:lk}) the ordering in time of the events
becomes crucial and instead of the situation in which the first $k$ waiting
time intervals have a different distribution, one can consider the situation
in which the last $k$ are characterized in a different way, corresponding to
\begin{eqnarray}
  \hat{\Lambda}_{\vec{k}} (u) & = & \hat{g}_{1} ( u- \mathcal{L} ) + \ldots +
  \hat{f}_{1} (u- \mathcal{L} ) \mathcal{E} \ldots \hat{f}_{n-1} (u-
  \mathcal{L} ) \mathcal{E} \hat{g}_{n} ( u- \mathcal{L} ) \nonumber\\
  &  & + \hat{f}_{1} (u- \mathcal{L} ) \mathcal{E} \ldots \hat{f}_{n} (u-
  \mathcal{L} ) \mathcal{E} \frac{1}{- \hat{f} (u- \mathcal{L} ) \mathcal{E}}
  \hat{g} ( u- \mathcal{L} ) \nosymbol . \label{eq:lkt} 
\end{eqnarray}
The index now denotes the inverse time ordering, from left to right, and one
can notice that $\hat{\Lambda}_{\vec{k}} (u) = \hat{\Lambda}_{\check{k}}^{T}
(u)$, where we have used the symbol $T$ to denote the inverse operator
ordering, i.e. $( A_{1} \ldots A_{n} )^{T} =A_{n} \ldots A_{1}$, since indeed
this evolution map can be obtained from Eq.~(\ref{eq:lk}) by inverting the
operator ordering. The two situations described by a modified quantum renewal
process together with a choice of operator ordering is schematically shown in
Fig.~\ref{fig:schema-lr}.

 The reference dynamics is now given by
\begin{eqnarray}
  \hat{\Lambda}_{\vec{0}} (u) & = & \frac{1}{\mathbbm{1} - \mathcal{E} \hat{f}
  (u- \mathcal{L} )} \hat{g} ( u- \mathcal{L} ) , \label{eq:l0r} 
\end{eqnarray}
with kernel
\begin{eqnarray}
  \widehat{\mathcal{K}}_{\vec{0}} (u) & = & \mathcal{L} + \hat{k} (u-
  \mathcal{L} ) ( \mathcal{E} - \mathbbm{1} ) , \label{eq:k0r} 
\end{eqnarray}
again connected to Eq.~(\ref{eq:l0l}) and Eq.~(\ref{eq:k0l}) respectively by
inverting the operator ordering. The master equation providing a closed
evolution equation for the dynamics given by Eq.~(\ref{eq:lkt}) can be written
as
\begin{eqnarray}
  \frac{\mathd}{\mathd t} \rho ( t ) & = & \! \int^{t}_{0} \mathd \tau
  \mathcal{K}_{\vec{k}} ( t- \tau ) \rho ( \tau ) , \label{eq:kkms} 
\end{eqnarray}
with a kernel simply given by $\widehat{\mathcal{K}}_{\vec{k}} (u) =
\mathcal{\widehat{\mathcal{K}}}^{T}_{\check{k}} ( u )$. The major difference
in considering as different the last $k$ waiting times is best appreciated
writing the master equation equivalent to Eq.~(\ref{eq:kkms}) but expressed
using the reference MK Eq.~(\ref{eq:k0r}) and a inhomogeneous contribution
\begin{eqnarray}
  \frac{\mathd}{\mathd t} \rho ( t ) & = & \! \int^{t}_{0} \mathd \tau
  \mathcal{K}_{\vec{0}} ( t- \tau ) \rho ( \tau ) + \mathcal{I}_{\vec{k}} ( t
  ) \rho ( 0 ) , \label{eq:wims} 
\end{eqnarray}
where now the inhomogeneous term takes the natural but involved expression
\begin{eqnarray}
  \mathcal{\widehat{\mathcal{I}}}_{\vec{k}} ( u ) & = &
  \hat{\Lambda}_{\vec{0}} (u)^{-1}
  \mathcal{\widehat{\mathcal{I}}}_{\check{k}}^{T} ( u )
  \hat{\Lambda}_{\vec{0}} (u) . \label{eq:ikr} \nonumber
\end{eqnarray}
At variance with the expression Eq.~(\ref{eq:ik}) appearing in the master
equation Eq.~(\ref{eq:msk}), where the effect of having a modified renewal
process is expressed as a simple correction in time, here the inhomogeneous
correction is obtained convoluting the time reversed inhomogeneous term with
free propagators forward and backward in time.

\section{Examples}\label{s6}

In order to exemplify the introduced formalism and to point out the different
dynamical behavior that can arise as a consequence of operator ordering, we
consider a few examples. The obtained class of legitimate MK, and therefore
CPT dynamics, depends both on the choice of jump transformations and
intermediate time evolution maps, as well as on the considered WTD
characterizing the different time intervals. Here we will focus in particular
on the comparison between a quantum renewal process and its modified
counterpart, as well as on the different dynamics arising by considering the
same sequence of events but in a different time operator ordering.

\begin{figure}[h]
 \tmtextbf{a)}
  {\includegraphics[width=0.4\textwidth]{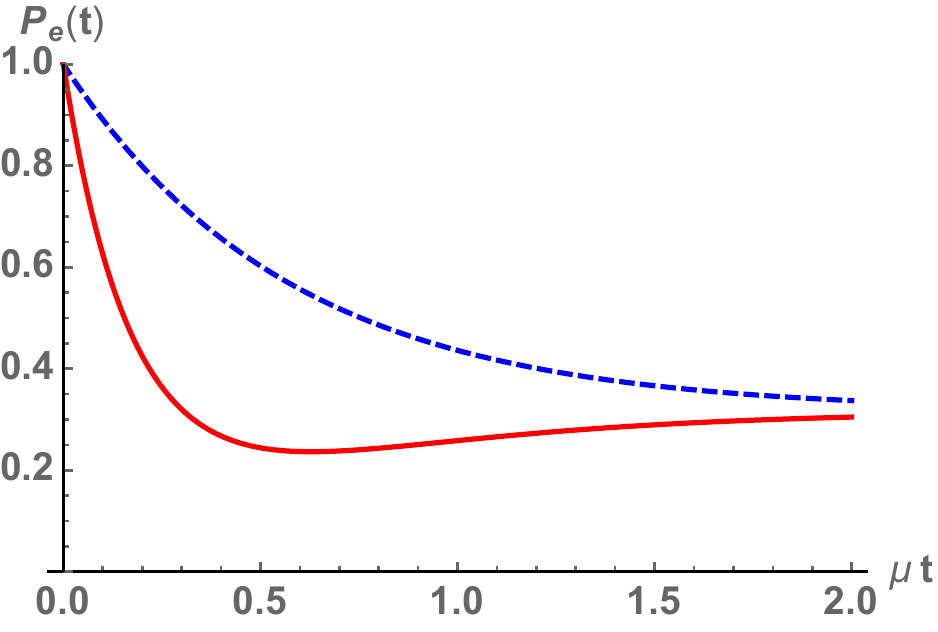}}
  \tmtextbf{b)}
  {\includegraphics[width=0.4\textwidth]{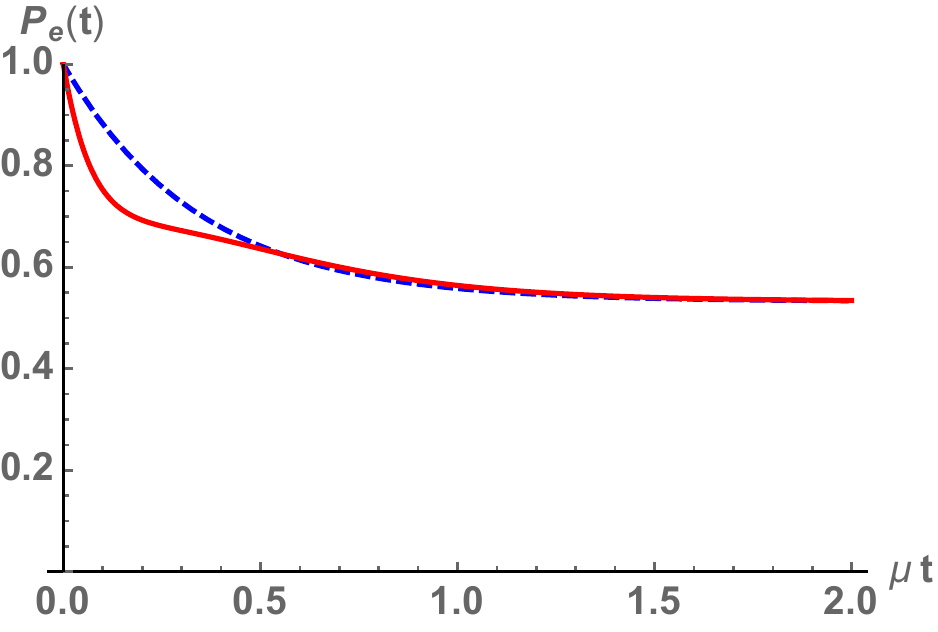}}
  \caption{\label{fig:cfr12}Behavior of the population of the excited state of
  a two-level system $P_{e} ( t )$. In each panel we compare a quantum renewal
  process and a modified version of it. Panel \tmtextbf{a)} dynamics described
  by an exponential damping with rate $\lambda_{3} =1.1$ in inverse units of
  time, intertwined with jumps described by a amplitude damping channel with
  parameter $\gamma =.8$ {\cite{Nielsen2000}}. The solid curve (red)
  corresponds to the modified quantum renewal process corresponding to
  Eq.~(\ref{eq:lk}), while the dashed curve (blue) to the unmodified one of
  Eq.~(\ref{eq:l0l}). Here $k=3$ so that for the modified process we have
  considered three exponential distributions with different rates in the ratio
  1:7:5. The populations are plotted as function of $\mu t$, with $\mu$ the
  parameter characterizing the first WTD. Panel \tmtextbf{b)} dynamics with
  the same damping rate and WTD in the ratio 1:5:10, but jumps described by
  the composition of a Pauli channel with Kraus operator $\sigma_{x}$ and an
  amplitude damping with parameter $\gamma =.43$.}
\end{figure}

Let us first consider the difference between Eq.~(\ref{eq:lk}), describing a
dynamics in which the first $k$ time intervals are characterized by a
different WTD, and its unmodified counterpart Eq.~(\ref{eq:l0l}). To this aim,
despite the fact that the obtained results are not constrained to finite
dimensional Hilbert spaces, we consider for the sake of simplicity a two-level
system. This allows in particular to have a simple matrix representation of
the different maps involved. Indeed, for a fixed basis of operators in the
Hilbert space, which we take to be given by the identity and the Pauli
matrices apart from a normalization factor, each map $\mathcal{A} [ \cdot ]$
can be represented by a four dimensional matrix with entries $A_{ij} =
\frac{1}{2} \tmop{Tr} ( \sigma_{i} \mathcal{A} [ \sigma_{j} ] )$, with
$i,j=0,1,2,3$. In this representation in particular map composition goes over
to matrix multiplication {\cite{Smirne2010b,Andersson2014a}}, so that
expressions of the form Eq.~(\ref{eq:lk}) and Eq.~(\ref{eq:l0l}) can be easily
evaluated. We take as reference dynamics a semigroup evolution describing
exponential dephasing and damping according to
\begin{eqnarray}
  \mathe^{\mathcal{L} t} [ \sigma_{i} ] & = & \mathe^{- \lambda_{i} t}
  \sigma_{i} \label{eq:semig} 
\end{eqnarray}
for $i=1,2,3$, while assuming an intermediate transformation $\mathcal{E}$
obtained by considering an amplitude damping channel, possibly composed with a
dephasing transformation. It is obviously crucial to consider non commuting
transformations, i.e. to fulfil the requirement $[ \mathcal{E} , \mathcal{L} ]
\neq 0$. According to Eq.~(\ref{eq:semig}) one can further exploit the
following matrix representation for maps given by functions of $u-
\mathcal{L}$
\begin{eqnarray}
  \hat{h} (u- \mathcal{L} ) & = & \tmop{diag} ( \hat{h} (u), \hat{h} (u+
  \lambda_{1} ), \hat{h} (u+ \lambda_{2} ), \hat{h} (u+ \lambda_{3} ) ) .
  \label{eq:wtdu} 
\end{eqnarray}
The final information necessary in order to fix the structure of the dynamical
map is given by the choice of WTD, which we take in the first instance as
exponential, i.e. of the form $\mu \mathe^{- \mu t}$, albeit with different
rates $\mu$. Such a WTD describes Poisson distributed events with rate $\mu$.
The difference in the obtained dynamics can be seen plotting the behavior in
time of the population of the excited state, as shown in Fig.~\ref{fig:cfr12}.
In particular it can be seen how the modified process can lead to a non
monotonic decrease of the population of the excited state.

\begin{figure}[h]
 \tmtextbf{a)}
  {\includegraphics[width=0.4\textwidth]{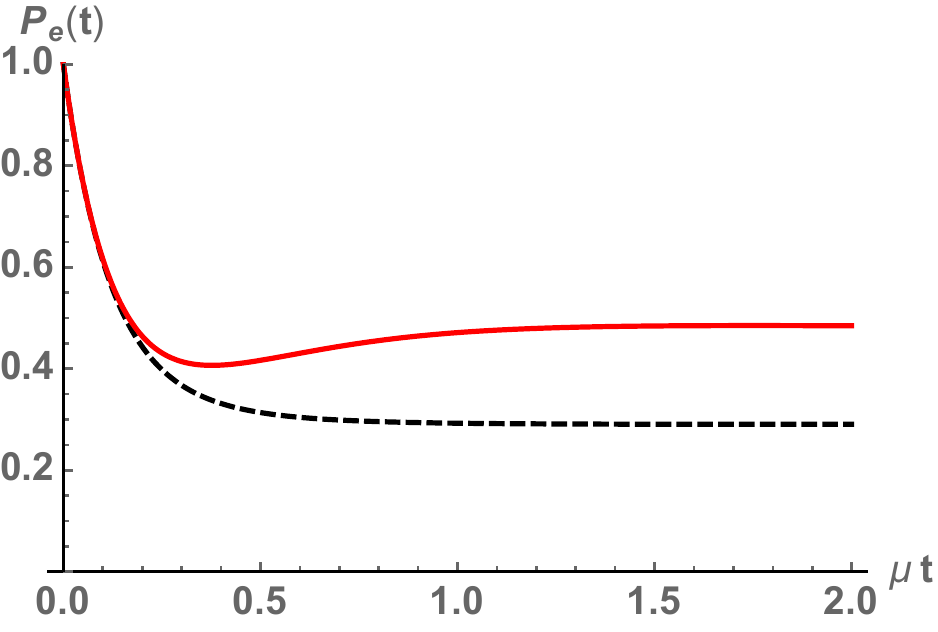}}
  \tmtextbf{b)}
  {\includegraphics[width=0.4\textwidth]{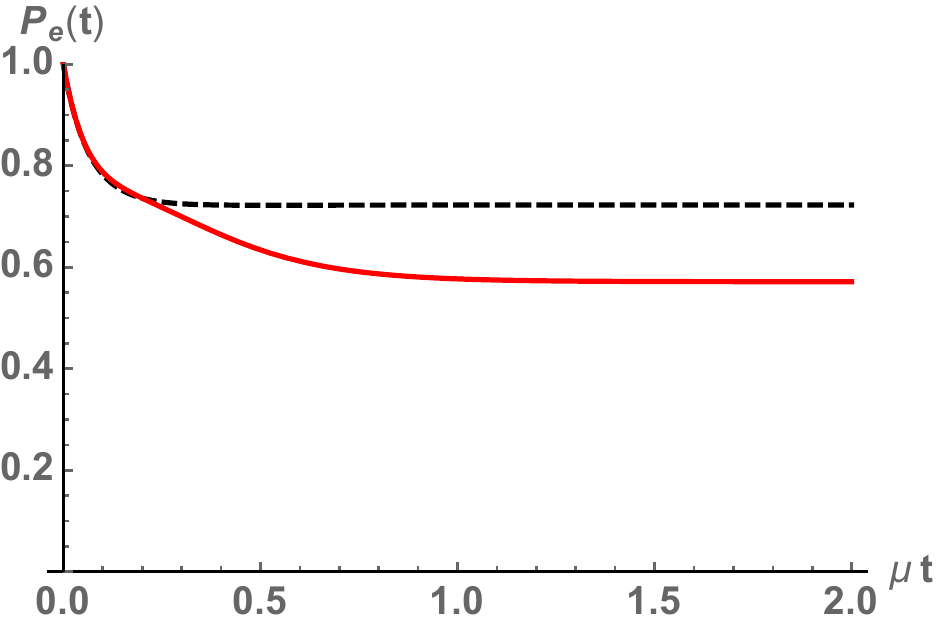}}
  \caption{\label{fig:cfr34}Time evolution of the occupation of the excited
  state $P_{e} ( t )$. The compared evolutions only differ for the time
  ordering. In both panels we consider a dynamics described by an exponential
  damping with rate $\lambda_{3} =4$ in inverse units of time. The process is
  characterized by four different exponential WTD, so that we have $k=4$. In
  panel \tmtextbf{a)} their rates are in the ratio 1:5:.1:.5, while in panel
  \tmtextbf{b)} the ratio is given by 1:10:20:15. Jumps are described by an
  amplitude damping channel with parameter $\gamma =.87$ {\cite{Nielsen2000}},
  further composed with a dephasing map in panel \tmtextbf{b)}. The solid
  curves (red) correspond to the quantum renewal process ordered as in
  Eq.~(\ref{eq:lk}), while the dashed curves (black) to the process with
  inverse time ordering as in Eq.~(\ref{eq:lkt}).}
\end{figure}

As a further illustration, we consider for the same system the situation in
which the dynamics only differ for the operator ordering. That is we consider
the distinct evolution maps Eq.~(\ref{eq:lk}) and Eq.~(\ref{eq:lkt}) for the
same specification of the generator $\mathcal{L}$ determining the intermediate
time evolution and the same channel describing the jumps, as well as WTD. To
this aim we still consider the semigroup dynamics given by
Eq.~(\ref{eq:semig}). The variety of possible different behavior is put into
evidence in Fig.~\ref{fig:cfr34}, where we have considered in the different
panels quantum processes only differing for the choice of channels and rates
of the involved waiting times. The inverse operator ordering corresponds to
the solid curves and typically brings in an important modification of the
dynamics before a stationary situation is reached.

The considered situations only provide an illustrative example of the
different behavior that can arise considering solutions of the quantum
dynamics that we have introduced as a quantum version of classical renewal
processes. Actual implementations will depend on dimensionality and details of
the considered \ system, as well as on the feature of the interactions
determining its reduced dynamics. It is important however to stress that
examples of realization of special cases in physical systems have already
appeared in the physical literature, e.g. modified WTD in the micromaser
dynamics {\cite{Herzog1995a,Cresser1996a,Cresser2019a}} or dynamics related to
MK corresponding to different orderings in the treatment of non-Markovian
collision models
{\cite{Giovannetti2012b,Ciccarello2013a,Lorenzo2016a,Lorenzo2017a}}.

\section{Conclusions and outlook}\label{s7}

We have constructed a new, large class of quantum MK, possibly including
inhomogeneous terms, which provide master equations whose solutions are indeed
CPT transformations. Though the construction of legitimate MK, providing more
general dynamics than the standard Lindblad one, has proven to be a very
difficult task {\cite{Barnett2001a,Shabani2005a,Campbell2012a}}, as we have
shown a natural and fruitful viewpoint is to make reference to classical
non-Markovian processes. In this framework we have considered a convenient
viewpoint for introducing a class of quantum transformations that have been
termed quantum renewal processes, due to the fact that they are built starting
from classical renewal processes. The basic ingredients of the construction
are indeed a collection of distributions over the time axis and a CPT channel
describing quantum transformations taking place in between an intermediate
semigroup evolution, after intervals dictated by the waiting time.

In the construction one can consider and put into evidence two important
variants. On the one hand, one can deal with modified processes, so that the
time intervals between subsequent quantum transformations are independent but
not identically distributed. On the other hand, for each legitimate MK one can
consider another distinct kernel, essentially obtained by transposition, still
leading to a well-defined dynamics and characterized by an inverted operator
ordering in the sequence of events. In all these dynamics a crucial role is
played by the typical quantum feature of non commutativity, bringing with
itself the relevant role played by operator ordering. Simple examples have
been provided, showing that indeed the interplay of these different features
can lead to a wide variety of behavior, further recalling that special cases
of this general framework have appeared in the description of physical systems
{\cite{Cresser1996a,Lorenzo2016a}}.

Despite significantly enlarging the known classes of MK leading to
well-defined reduced time evolutions, this contribution leaves open the
question about the most general characterization of such kernels. In
particular one might wonder what is the most general form of piecewise
dynamics leading to closed evolution equations in integral forms, and to what
extent these kind of dynamics can exhibit non-Markovian effects. These
questions naturally call for future investigations.

\section{Acknowledgments}

B.V. acknowledges support from the Joint Project ``Quantum Information
Processing in Non-Markovian Quantum Complex Systems'' funded by FRIAS,
University of Freiburg and IAR, Nagoya University, from the FFABR project of
MIUR and from the Unimi Transition Grant H2020.

\section{Appendix}\label{a1}

In order to obtain the expression Eq.~(\ref{eq:kerl}) for the kernel we start
from Eq.~(\ref{eq:wik}), which in Laplace transform leads to the following
general relationship between map and MK
\begin{eqnarray}
  \widehat{\mathcal{K}} ( u ) & = & u- \hat{\Lambda} ( u )^{-1} \nocomma ,
  \label{eq:ku} 
\end{eqnarray}
as well as the relationship
\begin{eqnarray}
  \frac{1}{\hat{g} ( u- \mathcal{M} )} & = & (u- \mathcal{M} ) + \hat{k} (u-
  \mathcal{M} ) , \nonumber
\end{eqnarray}
which follows from Eq.~(\ref{eq:kdef}) together with the Laplace transform
expression of the survival probability $u \hat{g} ( u ) = ( 1- \hat{f} ( u )
)$. Assuming now expression Eq.~(\ref{eq:ll}) for the time evolution map, with
a collection of intermediate time evolution maps $\{ \mathcal{F} ( t ) \}_{t
\in \mathbbm{R_{+}}}$ given by Eq.~(\ref{eq:expl}) together with $\mathcal{G}
( t ) = \mathe^{\mathcal{M} t}$, we come to the expression
\begin{eqnarray}
  \widehat{\mathcal{K}}_{l} ( u ) & = & u- ( \mathbbm{1} - \mathcal{E} \hat{f}
  (u- \mathcal{L} ) \mathcal{J} ) \hat{g} ( u- \mathcal{M} )^{-1} \nonumber\\
  & = & \mathcal{M} - \hat{k} (u- \mathcal{M} ) + ( \mathcal{E} \hat{f} (u-
  \mathcal{L} ) \mathcal{J} ) \hat{g} ( u- \mathcal{M} )^{-1} \nonumber\\
  & = & \mathcal{M} + ( \mathcal{E} \hat{f} (u- \mathcal{L} ) \mathcal{J} -
  \hat{f} (u- \mathcal{M} ) ) \hat{g} ( u- \mathcal{M} )^{-1} . \nonumber
\end{eqnarray}
For the case of exponential WTD $f ( t ) = \lambda \mathe^{- \lambda t}$, we
have the simple relationship $\hat{f} (u) = \lambda \hat{g} (u)$. In
particular this implies that if $\mathcal{J}$ becomes the trivial
transformation, $\mathcal{J} = \mathbbm{1}$, and the generators describing the
time evolution in the first time interval and the subsequent ones do coincide,
i.e. $\mathcal{M} = \mathcal{L}$, then we are left with
\begin{eqnarray}
  \widehat{\mathcal{K}} ( u ) & = & \mathcal{L} + ( \mathcal{E} \hat{f} (u-
  \mathcal{L} )- \hat{f} (u- \mathcal{L} ) ) \hat{g} ( u- \mathcal{L} )^{-1}
  \nonumber\\
  & = & \mathcal{L} + \lambda ( \mathcal{E} - \mathbbm{1} ) , \nonumber
\end{eqnarray}
and therefore in the time domain $\mathcal{K} ( t ) = \delta ( t ) [
\mathcal{L} + \lambda ( \mathcal{E} - \mathbbm{1} ) ]$, leading to the
memoryless evolution equation Eq.~(\ref{eq:split}).

\section{Appendix}\label{a2}

In order to prove the operator identity Eq.~(\ref{eq:fs}) we start from the
defining equation for the sprinkling distribution or renewal density
associated to a renewal process {\cite{Cox1965}}, namely
\begin{eqnarray}
  S ( t ) & = & f ( t ) + \int^{t}_{0} \mathd \tau f ( t- \tau ) S ( \tau ) ,
  \nonumber
\end{eqnarray}
leading in Laplace transform to the relation
\begin{eqnarray}
  \hat{S} (u) & = & \frac{\hat{f} (u)}{1- \hat{f} (u)} . \nonumber
\end{eqnarray}
Starting from the expression of the Laplace transform of the survival
probability we then have the following chain of operator identities
\begin{eqnarray}
  \hat{g} ( u- \mathcal{\mathcal{L}} ) \frac{1}{\mathbbm{1} - \mathcal{E}
  \hat{f} (u- \mathcal{L} )} & = & \frac{\mathbbm{1} - \hat{f} (u- \mathcal{L}
  )}{u- \mathcal{L}} \frac{1}{\mathbbm{1} - \mathcal{E} \hat{f} (u-
  \mathcal{L} )} \nonumber\\
  & = & \frac{\mathbbm{1} - \hat{f} (u- \mathcal{L} )}{u- \mathcal{L}} (
  \mathbbm{1} - \hat{f} (u- \mathcal{L} )- ( \mathcal{E} - \mathbbm{1} )
  \hat{f} (u- \mathcal{L} ) )^{-1} \nonumber\\
  & = & \frac{\mathbbm{1} - \hat{f} (u- \mathcal{L} )}{u- \mathcal{L}} \left(
  \left( \mathbbm{1} - ( \mathcal{E} - \mathbbm{1} ) \frac{\hat{f} (u-
  \mathcal{L} )}{\mathbbm{1} - \hat{f} (u- \mathcal{L} )} \right) (
  \mathbbm{1} - \hat{f} (u- \mathcal{L} ) ) \right)^{-1} \nonumber\\
  & = & \frac{1}{u- \mathcal{L}} \left( \mathbbm{1} - ( \mathcal{E} -
  \mathbbm{1} ) \frac{\hat{f} (u- \mathcal{L} )}{\mathbbm{1} - \hat{f} (u-
  \mathcal{L} )} \right)^{-1} \nonumber\\
  & = & \frac{1}{u- \mathcal{L}} \frac{1}{\mathbbm{1} - ( \mathcal{E} -
  \mathbbm{1} ) \hat{S} (u- \mathcal{L} )} , \nonumber
\end{eqnarray}
leading to Eq.~(\ref{eq:fs}).

\section{Appendix}\label{a3}

We now prove that the master equation, for the case of a modified quantum
renewal process with the first $k$ time intervals following a different
distribution, can be expressed making reference to the MK for the unmodified
case together with inhomogeneous contributions of the form Eq.~(\ref{eq:ik}).
To this aim let us start from the general expression of MK master equation
with inhomogeneous term given by Eq.~(\ref{eq:l}), which in Laplace transform
reads
\begin{eqnarray}
  \hat{\Lambda} (u) & = & ( u- \widehat{\mathcal{W}} ( u ) )^{-1} (
  \mathbbm{1} + \mathcal{\widehat{\mathcal{I}}} ( u ) ) . \label{eq:ilt} 
\end{eqnarray}
We now want to identify the inhomogeneous term for an evolution map given by
Eq.~(\ref{eq:lk}), taking as reference kernel
$\widehat{\mathcal{K}}_{\check{0}} (u)$ as in Eq.~(\ref{eq:k0l}), so that $(
u- \widehat{\mathcal{K}}_{\check{0}} (u) )^{-1}$ according to
Eq.~(\ref{eq:ku}) corresponds to the unmodified dynamics Eq.~(\ref{eq:l0l})
and we are left with
\begin{eqnarray}
  \hat{\Lambda}_{k} (u) & = & \hat{g} ( u- \mathcal{L} ) \frac{1}{\mathbbm{1}
  - \mathcal{E} \hat{f} (u- \mathcal{L} )} ( \mathbbm{1} +
  \mathcal{\widehat{\mathcal{I}}}_{k} ( u ) ) . \label{eq:lii} 
\end{eqnarray}
Considering in the first instance $k=1$ we obtain the identity
\begin{eqnarray}
  \hat{g}_{1} ( u- \mathcal{L} ) + \hat{g} ( u- \mathcal{L} )
  \frac{1}{\mathbbm{1} - \mathcal{E} \hat{f} (u- \mathcal{L} )} \mathcal{E}
  \hat{f}_{1} (u- \mathcal{L} ) & = & \hat{g} ( u- \mathcal{L} )
  \frac{1}{\mathbbm{1} - \mathcal{E} \hat{f} (u- \mathcal{L} )} ( \mathbbm{1}
  + \mathcal{\widehat{\mathcal{I}}}_{1} ( u ) ) . \label{eq:inter} 
\end{eqnarray}
We now recall that $u \hat{g}_{1} ( u ) = ( 1- \hat{f}_{1} ( u ) )$, while the
sprinkling distribution for a modified process obeys
\begin{eqnarray}
  S_{1} ( t ) & = & f_{1} ( t ) + \int^{t}_{0} \mathd \tau f ( t- \tau ) S_{1}
  ( \tau ) , \nonumber
\end{eqnarray}
so that
\begin{eqnarray}
  \hat{S}_{1} (u) & = & \frac{\hat{f}_{1} (u)}{1- \hat{f} (u)} . \label{eq:s1}
\end{eqnarray}
We have in particular the relation
\begin{eqnarray}
  \hat{g}_{1} ( u- \mathcal{L} ) & = & \hat{g} ( u- \mathcal{L} ) (
  \mathbbm{1} + \hat{S} (u- \mathcal{L} )- \hat{S}_{1} (u- \mathcal{L} ) ) ,
  \nonumber
\end{eqnarray}
allowing to write the r.h.s. of Eq.~(\ref{eq:inter}) in the form
\[ \hat{g} ( u- \mathcal{L} ) \left[ \mathbbm{1} + \hat{S} (u- \mathcal{L} )-
   \hat{S}_{1} (u- \mathcal{L} )+ \frac{1}{\mathbbm{1} - \mathcal{E} \hat{f}
   (u- \mathcal{L} )} \mathcal{E} \hat{f}_{1} (u- \mathcal{L} ) \right] , \]
so that Eq.~(\ref{eq:inter}) becomes equivalent to
\begin{eqnarray}
  ( \mathbbm{1} - \mathcal{E} \hat{f} (u- \mathcal{L} ) ) \left( \mathbbm{1} +
  \hat{S} (u- \mathcal{L} )- \hat{S}_{1} (u- \mathcal{L} )+
  \frac{1}{\mathbbm{1} - \mathcal{E} \hat{f} (u- \mathcal{L} )} \mathcal{E}
  \hat{f}_{1} (u- \mathcal{L} ) \right) & = & ( \mathbbm{1} +
  \mathcal{\widehat{\mathcal{I}}}_{1} ( u ) ) \nosymbol . \nonumber
\end{eqnarray}
This in turn leads to
\begin{eqnarray}
  \mathcal{\widehat{\mathcal{I}}}_{1} ( u ) & = & ( \mathbbm{1} - \mathcal{E}
  \hat{f} (u- \mathcal{L} ) ) ( \hat{S} (u- \mathcal{L} )- \hat{S}_{1} (u-
  \mathcal{L} ) ) + \mathcal{E} ( \hat{f}_{1} (u- \mathcal{L} )- \hat{f} (u-
  \mathcal{L} ) ) \nonumber
\end{eqnarray}
but according to Eq.~(\ref{eq:s1}) we have
\begin{eqnarray}
  \hat{f}_{1} (u- \mathcal{L} ) - \hat{f} (u- \mathcal{L} ) & = & (
  \mathbbm{1} - \hat{f} (u- \mathcal{L} ) ) ( \hat{S}_{1} (u- \mathcal{L} )-
  \hat{S} (u- \mathcal{L} ) ) \label{eq:f1fs} 
\end{eqnarray}
and therefore finally
\begin{eqnarray}
  \mathcal{\widehat{\mathcal{I}}}_{1} ( u ) & = & ( \mathcal{E} - \mathbbm{1}
  ) ( \hat{S}_{1} (u- \mathcal{L} )- \hat{S} (u- \mathcal{L} ) ) . \nonumber
\end{eqnarray}
To prove the relation in the general case we proceed by induction, omitting
the common argument $u- \mathcal{L}$. We consider the case $k+1$, which due to
the expression of the evolution map corresponds to
\begin{eqnarray}
  ( \mathbbm{1} - \mathcal{E} \hat{f} ) ( \mathbbm{1} + \hat{S} - \hat{S}_{1}
  ) + \ldots + ( \mathbbm{1} - \mathcal{E} \hat{f} ) ( \mathbbm{1} + \hat{S} -
  \hat{S}_{k+1} ) \mathcal{E} \hat{f}_{k} \ldots \mathcal{E} \hat{f}_{1} +
  \mathcal{E} \hat{f}_{k+1} \ldots \mathcal{E} \hat{f}_{1} & = & \mathbbm{1} +
  \mathcal{\widehat{\mathcal{I}}}_{k+1} ( u ) \nonumber
\end{eqnarray}
but assuming assume $\mathcal{\widehat{\mathcal{I}}}_{k} ( u )$ to be of the
form Eq.~(\ref{eq:ik}), adding and subtracting the term $\mathcal{E}
\hat{f}_{k} \ldots \mathcal{E} \hat{f}_{1}$ we are left with
\begin{eqnarray}
  \mathcal{\widehat{\mathcal{I}}}_{k+1} ( u ) & = &
  \mathcal{\widehat{\mathcal{I}}}_{k} ( u ) + [ ( \mathbbm{1} - \mathcal{E}
  \hat{f} ) ( \mathbbm{1} + \hat{S} - \hat{S}_{k+1} ) - \mathbbm{1} +
  \mathcal{E} \hat{f}_{k+1} ] \mathcal{E} \hat{f}_{k} \ldots \mathcal{E}
  \hat{f}_{1} \nonumber\\
  & = & \mathcal{\widehat{\mathcal{I}}}_{k} ( u ) + [ ( \hat{S} -
  \hat{S}_{k+1} ) - \mathcal{E} \hat{f} ( \hat{S} - \hat{S}_{k+1} ) +
  \mathcal{E} ( \hat{f}_{k+1} - \hat{f} ) ] \mathcal{E} \hat{f}_{k} \ldots
  \mathcal{E} \hat{f}_{1} \nonumber\\
  & = & \mathcal{\widehat{\mathcal{I}}}_{k} ( u ) + ( \mathcal{E} -
  \mathbbm{1} ) ( \hat{S}_{k+1} - \hat{S} ) \mathcal{E} \hat{f}_{k} \ldots
  \mathcal{E} \hat{f}_{1} , \nonumber
\end{eqnarray}
where we have used again Eq.~(\ref{eq:f1fs}), thus obtaining the general
result.

\end{document}